\documentclass[12pt,preprint]{aastex}
\usepackage{psfig}
\newcommand{\beq}{\begin{equation}}
\newcommand{\eeq}{\end{equation}}
\newcommand{\beqn}{\begin{eqnarray}}
\newcommand{\eeqn}{\end{eqnarray}}

\newcommand{\no}{\noindent}
     
\begin{document}

\title{\bf{The Galactic Mergers and Gravitational Unbound Populations}}

\author{Yu-Ting Wu and Ing-Guey Jiang}

\affil{
{Department of Physics and Institute of Astronomy,}\\ 
{National Tsing-Hua University, Hsin-Chu, Taiwan} 
}

\begin{abstract}
Motivated by the observations on the intra-cluster light (ICL) and 
inter-galactic stellar populations,
n-body simulations are used to model 
the galactic merging events
as a goal to investigate the production and distribution 
of gravitational unbound populations (GUPs).
Both the parabolic and hyperbolic mergers are considered and each category
includes six models with different relative orientations between two galaxies.
Our results show that there are more (about a factor of two)
GUP after a hyperbolic merging event 
than after a parabolic one.
In general, depending on 
the relative orientation and also the relative velocity 
of the two galaxies in a merging pair, 
a head-on collision of a galaxy pair 
would only make a tiny fraction (less than one percent)
of the initial stellar mass become luminous GUP but 
a considerable fraction (eight to fourteen percent) 
of the dark matter become dark GUP. 


\end{abstract}

\noindent
{\bf Key words:} galaxies: interactions; galaxies: stellar content;
galaxies: dynamics

\section{The Introduction and Motivation}

The origin of the diffused light between galaxies, 
as first discovered by Zwicky (1951) in the Coma cluster,
has been a long term puzzle because in the standard picture, 
stars in the universe are 
formed and located within galaxies. However, Zwicky (1951)'s result implies 
that there are many luminous stars 
located between galaxies. 
A significant amount 
of stars must be far from galaxies, as implied by the
brightness of the diffused light.

In order to further investigate this problem,
over the years, there have been  
many more observations made about these inter-galactic
objects. 
Red giants have been  
observed in a blank field in the Virgo Cluster as shown
in Durrell et al. (2002). 
Planetary nebulae have been observed in the inter-galactic space
of the M81 groups of galaxies
(Feldmeier et al. 2004), and also in the Virgo Cluster  
(Ferguson et al. 1998, Arnaboldi et al. 2002)
and Coma Cluster (Gerhard et al. 2005).
Moreover, Gal-Yam et al. (2003) even found
supernovae between galaxies.

Among these works, Durrell et al. (2002)
suggested that intra-cluster stars could contribute $10\%\sim20\%$ 
of the total I-band luminosity of the Virgo cluster.  Furthermore, 
Mihos et al. (2005) recently observed 
Virgo cluster's core, as shown in their Figure 1 and 2, and found
that there are a lot of intra-cluster stars and tidal 
tails between galaxies in the cluster.

With the above observational results, 
many interesting questions on the production of inter-galactic populations 
could be asked. 
Did these stars form through the collapse of inter-galactic medium ? 
Are they the remnants of stellar groups during the early phase of the 
formation of a cluster of galaxies ? 
Are they the outcome of continuous stripping due to the
merging and interaction between galaxies ? 
In order to obtain good answers, many theoretical studies on this subject 
have been done.
For example, 
focusing on clusters of galaxies,
cosmological n-body simulations have been employed to investigate
the origins of intra-cluster populations (Please see
Napolitano et al. 2003, Murante et al. 2004,
Willman et al. 2004, Sommer-Larsen et al. 2005, Rudick et al. 2006, 
Murante et al. 2007).
Using an observational definition of intra-cluster light (ICL)
to be luminosity at a surface brightness $\mu_V > 26.5$ mag 
$ {\rm arcsec}^{-2}$, Rudick et al. (2006) found that about 
$10\%\sim15\%$ of the clusters' luminosity is at ICL surface brightness.
In their simulations, 
the tidal stripping of cluster galaxies 
could be one of the main mechanisms to produce ICL. However, 
Murante et al. (2007) concluded that the majority of the 
diffused stellar component is produced during the merging in the formation
history of the bright cluster galaxies. 

Although inter-galactic stellar populations and the 
ICL have been investigated by many groups both
observationally and theoretically, as reviewed in Zibetti (2008),
one of the main issues is about the very definition of 
inter-galactic stellar populations and the ICL. All photometric studies 
have to define criteria, such as surface brightness thresholds etc.
to isolate the light of inter-galactic stars from the galactic signal.
Note that the previously mentioned work on clusters of galaxies
employed different 
working definitions of the ICL with their own chosen criteria.
However, from the dynamical point of view,  
the light from the stars which are not bound to any galactic
potential (with dark matter included) seems to be a fundamental choice.

On the other hand, due to that
van Dokkum et al. (1999) presented many examples of mergers of 
E/S0, or early-type spiral galaxies, 
Stanghellini et al.(2006) went on to study the
production of inter-galactic populations through the mergers 
of elliptical galaxies. 
In their work, a pair of elliptical galaxies 
with a given initial mass ratio and initial relative velocity,
represented by 
spherical distributed N-body systems, is set to 
have a head-on collision.
The massive halo of dark matter is not considered in the above simulations.
They conclude that this scenario is helpful to feed $5\% \sim 20\%$ 
of the total initial mass as unbound stars.

The work in Stanghellini et al.(2006) presents a good example
to study the production of gravitational unbound population (GUP)
by a merging event. This types of work could compromise the cosmological
n-body simulations on clusters, in the sense that (a) the detail 
relationship between GUP productions and the ways of merging could 
be investigated and (b) the members of GUP are well defined and easy
to be identified in simulations.
As Zibetti (2008) mentioned, in one particular system of galaxies, 
GUP shall be different 
from those photometric ICL, and to identify
both of them observationally would be an important future work.

To take a further step from Stanghellini et al.(2006),
we here study the merging events of spiral galaxies.
The discs are used to represent the stellar
components of spiral galaxies and there is no gaseous component. 
Thus, our model galaxies
may be regarded as 
early-type spiral galaxies  
or late-type spiral galaxies in which the gaseous effect is ignored. 
Although ICL is our motivation for this work, we will only investigate
the production of GUP here.
The well-defined GUP could give strong constraint on the 
dynamical evolution of mergers. 
 
Moreover, 
because dark matter is more massive than the stellar part 
(Faber 1987) and shall be dynamically important,  
we include the dark haloes in the systems.
Indeed, the dark halo plays a crucial role in galactic dynamics. 
For example, Jiang \& Binney (1999) demonstrated that the infall 
of dark matter could influence the orientation of the dark halo,
and then warp the stellar disc. Jiang \& Binney (2000) further showed
that the dynamical friction from the Milky Way's dark halo could 
successfully explain the orbit of the Sagittarius dwarf galaxy.
 




In the following, we describe the details of our models in \S 2
and present the results of our N-body simulations on mergers 
in \S 3. The conclusions are in \S 4.

\section{The Model}

The merging of disc galaxies is one of the most important subjects in
galactic dynamics. A well-known example of merging is the interacting galaxies,
NGC 4038/4039 pair, and the corresponding numerical simulations 
in Toomre and Toomre (1972). The improvement of computing facilities 
triggers many further investigations. 
Dubinski et al.(1996) presented an interesting 
study about the relation between the mass of dark haloes and
the length of tidal tails in merging galaxies.
Springel \& White (1999) investigated the tidal tails 
in pairs of merging disc galaxies, from the view of 
cold dark matter (CDM) cosmologies.
Naab, Burkert \& Hernquist (1999) used the merging of disc galaxies to
explain the origins of boxy and disky elliptical galaxies.	
Naab \& Burkert (2003) performed a huge number of N-body simulations
on the binary mergers of disk galaxies, in order to do a large parameter
survey and obtain the general properties.  
Bournaud et al. (2007) investigated the role of multiple minor mergers 
from the point of galaxy formation.
In addition to the above, please also see more recent work in 
Lotz et al. (2008),
Johansson et al. (2009),
Thomas et al. (2009), 
and Naab \& Ostriker (2009).

In this paper, we study the head-on mergers of spiral 
galaxies and their contributions on the production of GUP. 
A pair of galaxies with dark haloes will be employed for the
simulations of merging events. 
Each galaxy includes two components, the stellar disc and the dark halo, 
with the density profiles as 
described in Hernquist (1993).
These two components will be allowed to influence each other and 
relax to approach a new combined equilibrium.
The parallelized version of the code, GADGET, is used for 
all N-body simulations (Springel et al. 2001),
and the force softening is set as 0.075 kpc for both 
the stellar disc and dark halo.

\subsection{The Initial Profiles}
The density structure of the stellar part is 
\begin{equation}
\rho_{d}(R,z)=\frac{M_d}{4 \pi h^2 z_0} \exp(-R/h) {\rm sech}^2(\frac{z}{z_0}),
\label {eq2-1}
\end{equation}
 where $M_d$ is the disc mass, 
$h$ is the radial scale length, and $z_0$ is a
vertical scale length. Thus, the above function
determines the initial positions of the disc particles.

On the other hand, the halo density profile is
\begin{equation}
\rho_{h}(r)= 
\frac{M_h}{2 \pi ^{3/2}}\frac{\alpha}{r_t r^2_c}\frac{\exp(-r^2 / r
^2 _t)}{\frac{r^2}{r^2_c}+1} ,
\label {eq2-4}
\end{equation}
where $M_h$ is the halo mass, $r_t$ is the tidal radius, 
$r_c$ is the core radius, and $\alpha$ is defined as:
\begin{equation}
\alpha = \{1-\sqrt{\pi}q \exp(q^2)[1-{\rm erf}(q)]\}^{-1},
\label {eq2-5}
\end{equation}
where $q = r_c / r_t$ and ${\rm erf}(q)$ 
is the error function as a function of $q$.
The halo particles' initial positions
are given following the above profile.

To assign initial velocities on the disc and halo particles, 
a rigorous procedure is used to calculate the corresponding phase-space 
distribution functions from the given
density profiles. For spherical systems, Eddington's formula
(Binney \& Tremaine 1987) can be used for this purpose. 
Thus, as in Binney et al. (1998), 
we numerically calculate the halo's phase-space distribution function 
from $\rho_{h}(r)$ and determine the halo particles' initial velocities
according to this distribution function.
For non-spherical systems, such as the stellar disc, there is no 
analytical formula to be used to obtain the phase-space distribution 
function directly from the density profile. We thus have to determine
the disc particles' velocities from the moments of the 
collision-less Boltzmann equation as in 
Hernquist (1993).
     

\subsection{The Units}

In our simulations, the unit of length is 1 kpc, 
the unit of mass is $10^{10} M_\odot$, 
the unit of time is $9.8\times10^8$ years, and  
the gravitational constant $G$ is 43007.1.

\subsection{The Disc-Halo Equilibrium}

There are two components in our galaxy, i.e. the stellar disc and the dark halo. 
Using the above units, 
the parameters for the galaxy are set as follows: 
for the dark halo, 
$M_h=32.48, r_c=3.5, r_t=35.0$, and the number of halo particles $N_h=58000$;
for the stellar disc,
$M_d=5.6, h=3.5, z_0=0.7$, and the number of disc particles $N_d=10000$.
Because a test particle's velocity at disc's half-mass radius, $R_{1/2}=5.95$, 
is $v_{1/2}=214.77$, the dynamical time is 
defined to be $T_{dyn} \equiv 2 \pi R_{1/2}/ v_{1/2}=0.174$, 
which is $1.7\times10^8$ years.


After the construction of the N-body models of the dark halo 
and the stellar disc, 
we combine these two components to set up a disc-halo system.
This system is allowed to relax to approach a new equilibrium.
The total kinetic energy $K$ and the total potential energy $U$
are calculated, and the value of ${2K}/{|U|}$ should be around one
when the disc-halo system is in equilibrium, according
to the virial theorem. We call ${2K}/{|U|}$ the 
{\it virial ratio} from  hereafter.

In Fig. 1, the virial ratio as a function of time is plotted
in the top panel, and the radius enclosing 
10\%, 25\%, 50\%, 75\%, 90\% of the total mass, 
i.e. the Lagrangian radii, as a function of time
are shown in the bottom panel.
They show that the system approaches equilibrium 
at $t=10 T_{dyn}$.
In addition, the variation of total energy is $0.013\% $, giving 
a satisfactory energy conservation.
Thus, the disc-halo system at $t=10 T_{dyn}$ of relaxation
will be used to represent one of the galaxies at the beginning 
of a merging episode.

\section{The N-Body Simulations}

For the simulations on the merger of a pair of galaxies, the parameter
space is too big to be completely explored in one paper. 
The initial mass ratio of the pair is fixed to be ${M_2}/{M_1}=1$,
with an initial separation $r_i=300$ kpc and an impact parameter $b=0$. 
Based on these parameters, the orbital energy of a merger is defined as:
\begin{equation}
E_{orb} \equiv \frac{1}{2}\frac{M_1M_2}{M_1+M_2}v^2 _i-\frac{GM_1 M_2}{r_i},
\end{equation}
where $v_i$ is the initial relative velocity.

On the other hand,
as shown in van Dokkum et al. (1999), there are many different possible 
relative angles between two discs in a merger.
In order to study this effect, we choose six different relative 
orientations between the angular momentums of two discs.
The galaxy located at $(x, y, z)=(0, -150, 0)$ would always have 
an angular momentum with the direction $\hat L$=(0, 0, 1), 
but the galaxy at $(x, y, z)=(0, 150, 0)$ would have an angular momentum
with six different directions.
For parabolic (hyperbolic) mergers, the galaxy at $(x, y, z)=(0, 150, 0)$ 
would have an angular momentum with direction 
$\hat {L_1}$=(0, 0, 1) in Model P1 (H1),
$\hat {L_2}$=(0, -1, 0) in Model P2 (H2),
$\hat {L_3}$=(0, 0, -1) in Model P3 (H3),
$\hat {L_4}$=(0, 1, 0) in Model P4 (H4),
$\hat {L_5}$=(1, 0, 0) in Model P5 (H5),
$\hat {L_6}$=(-1, 0, 0) in Model P6 (H6), respectively.
The discs of two galaxies are said to be parallel in Model P1 (H1), 
anti-parallel in Model P3 (H3), and perpendicular in 
the rest models.
 
\subsection{The Parabolic Mergers}

For the parabolic mergers, the initial orbital energy $E_{orb}=0$.
Thus, the two galaxies of the parabolic mergers have an 
initial relative velocity $v_i=147.8$ km/s. 

\subsubsection{Evolution}

Fig. 2(a) shows the virial ratio, $2K/|U|$, 
of the galaxy pair (with dark haloes) 
as a function of time for Model P1  
during t=$0\sim 39 T_{dyn}$ ($T_{dyn}$ 
is the dynamical time). 
There are strong interactions between two galaxies during 
t=$5 \sim 15 T_{dyn}$, so that the virial ratio goes away from one.
The first peak of $2K/|U|$ is around $t=8 T_{dyn}$ and the 2nd peak
is around $t=10 T_{dyn}$. The minimum of $2K/|U|$ is around $t=9 T_{dyn}$.

On the other hand, Fig. 2(b) shows the distance between two stellar
components of the galaxies during the merging. 
We plot the center of mass 
of each stellar component as a function of time in Model P1.
The location of each stellar component in this plot is 
defined by the center of mass 
of its particles. Indeed, the first close encounter
happens at $t=8 T_{dyn}$ and the 2nd one is at $t=10 T_{dyn} $. 
In between these two encounters,
the stellar components of two galaxies are well separated at $t=9 T_{dyn}$. 
After $t=10 T_{dyn}$, two galaxies are merged and a merger has formed.

The initial relative orientation of discs in a merger does not affect the
timing of principle dynamical processes, so that the corresponding
plots of other parabolic models are similar with Fig. 2 and not shown
here. 
However, the particle distributions of the stellar components in the mergers
are different in these models as shown in Fig. 3. 
Using Model P1 as an example,
Fig. 4 shows the distribution of stellar particles on the $x-y$ plane 
at different times.  
Initially, at $t=0$, two discs are well separated by 300 kpc. 
They come a bit closer
at $t=4 T_{dyn}$ and merge at $t=8 T_{dyn}$.  
After a violent mixing process from $t=9$ to $10 T_{dyn}$, 
the two systems combine to approach a new equilibrium after t=$15$,
and some particles escape far from the central part, as shown at
$t=27$ and $39 T_{dyn}$. 

The dark-matter-particle distributions in the mergers
of Model P1-P6 are shown in Fig. 5. 
Fig. 6 shows the distribution of dark halo particles in Model P1.
At $t=4 T_{dyn}$, the two haloes are very close and almost touching.
They are merged between $t=8$ and $9 T_{dyn}$, 
expanding further from $t=10$ to $15 T_{dyn}$. The outer part
expands even further, causing some particles to escape from the system. 

In Fig.3-6, the blue dots represent the unbound particles which
were belong to the galaxy initially located at $(x,y,z)=(0, -150, 0)$,
and the pink dots represent the unbound particles which
were belong to another galaxy.
 
\subsubsection{The Density Profiles}

In order to understand the distribution of the stellar
particles of the merger, 
we calculate the mass density of stellar particles 
as a function of $r$, where $r$ is the distance from 
the center of mass of all stellar particles.
Fig. 7(a) gives the density profiles of the stellar components in
the models of the parabolic mergers 
at $t=39 T_{dyn}$. 
It shows that the central density yielded from Model P1 
is higher than that which is 
yielded by Model P2-P6.
The central density given by Model P3 is very low, indicating that 
the whole system is stretched out.
The central densities given by other models are in between.

Fig. 7(b) 
shows the percentage of accumulated  stellar mass as a function of $r$.
For the region with $r<3$, the solid curve (Model P1)
is at the top and the short-dashed curve (Model P3) is at the bottom, 
which is consistent with
the results shown in Fig. 7(a).
However, near $r=20$, all the curves approach similar values.

On the other hand, 
Fig. 7(c) and 7(d) give the density profiles and 
the percentage of accumulated mass for the dark matter. 
They show that the differences between models are smaller than
the corresponding ones for the stellar components in Fig. 7(a)-(b).
 


\subsubsection{The Gravitational Unbound Population}

  
We calculate the total energy of both 
stellar and dark-matter particles, in order to identify the GUP,
and then determine their fractions. 
In Table 1, the first row gives the percentage of stellar GUP 
particles (compared to the total number of stellar particles) and
the 2nd row lists the percentage of dark-matter GUP 
particles 
(compared to the total number of dark-matter particles) 
at $t=39 T_{dyn}$ in parabolic mergers.
For the stellar GUP,
the average percentage is $m_{Ps}=0.1583$, and the standard deviation is 
$\sigma_{Ps}=0.0524$. Almost all values are within or near the boundary
of the interval 
$[m_{Ps}-\sigma_{Ps},m_{Ps}+\sigma_{Ps}]=[0.1059, 0.2107]$. 
For the dark-matter GUP, the mean is $m_{Pd}=7.7883$
with a standard deviation $\sigma_{Pd}=0.0508$. The interval
$[m_{Pd}-\sigma_{Pd},m_{Pd}+\sigma_{Pd}]=[7.7375,7.8391]$
Only the dark-matter GUP of Model P6 is a bit smaller and has a value
about  $m_{Pd}-2\sigma_{Pd}$.

\begin{table}[h]
\begin{center}
\begin{tabular}{|l||c|c|c|c|c|c|} \hline
{Model} & P1 & P2 & P3 & P4 & P5 & P6\\
\hline
\hline
stellar GUP (\%) & 0.16 & 0.10 & 0.18 & 0.08 & 0.22 & 0.21 \\ \hline
dark-matter GUP (\%) & 7.84 & 7.80 & 7.83 & 7.76 & 7.81 & 7.69 \\ \hline 
\end{tabular}
\caption[The GUP of Model P1-6]{The percentages of 
stellar GUP and dark-matter GUP in parabolic mergers.
}
\end{center}
\end{table} 

\subsection{The Hyperbolic Mergers}

In order to examine the effects of larger relative velocities in a merging
event, we here investigate hyperbolic mergers, 
in which the orbital energy $E_{orb}>0$.
We model six hyperbolic mergers with different relative orientations.
All the parameters of the hyperbolic mergers are 
the same as in the parabolic mergers, 
except that we assume the initial relative velocities of merging galaxies 
in these models (Model H1-H6) to be
$v_{max}$, which can be calculated from (Binney \& Tremaine 1987) 
\begin{equation}
v_{max} = 1.2 \sqrt{<v^2>},
\end{equation}
where $<v^2>$ is the mean-square velocity of all particles
in one galaxy (both stellar and dark components are included).
If the relative velocity between two galaxies is larger than 
$v_{max}$, it is likely that they will penetrate through each other,
and cannot combine to be a merger. 
Using $v_{max}$ as the initial relative velocity would highlight the
effects of the larger relative merging velocity.
We find that $v_{max}$ is equal to $273.1$ km/s for our system.


Using the model with anti-parallel discs as an example,
Fig. 8(a)
shows the virial ratio, $2K/|U|$, of one merging system (including both 
stellar and dark components) as a function of time 
for Model H3 during t=$0\sim 39 T_{dyn}$. 
There are direct collisions and strong interactions between two galaxies, 
so that the virial ratio $2K/|U|$ goes far from 1.
It shows that the first peak of $2K/|U|$ is about $t=5 T_{dyn}$ 
and the 2nd peak
is about $t=10 T_{dyn}$. The minimum of $2K/|U|$ is about $t=8 T_{dyn}$.

Moreover, Fig. 8(b) shows the distance between two stellar
components of the galaxies during the merging of Model H3. 
The first close encounter
happens at $t=5 T_{dyn}$, and then two galaxies move away from each other
until $t=8 T_{dyn}$, at which they are separated for more than 60 kpc.
They collide again at $t=10 T_{dyn} $ and get mixed together afterward.
The corresponding plots of other hyperbolic mergers are similar with the above,
thus are not shown in the paper.
The distributions of stellar particles on the $x-y$ plane  
for Model H1-H6 are shown in Fig. 9.
Using Model H3 (Fig. 10) as an example,
at $t=0$, two discs are well separated by 300 kpc. They get closer
at $t=2 T_{dyn}$ and soon merge at $t=5 T_{dyn}$.  
The violent mixing processes from $t=8$ to $10 T_{dyn}$ produce a complicated
structure, due to that the spins of two discs are completely 
opposite initially.
The system approaches equilibrium gradually 
and some particles escape far from the central part of the merger, 
as shown at
$t=15$, $27$, and $39 T_{dyn}$. 
The dark-matter-particle distributions in the mergers
of Model H1-H6 at $t=39 T_{dyn}$ are shown in Fig. 11
and the distributions of dark halo particles of Model H3 
at different times  are shown in Fig. 12.
The halo expands slightly larger than the ones in the parabolic mergers.


Here we also 
calculate the mass density of stellar particles in each spherical shell
and obtain the density of the stellar component as a function of $r$.
Fig. 13(a) 
shows the density profiles of stellar components in
Model H1-H6 at $t=39 T_{dyn}$. 
It shows that the central density of Model H1 is higher than 
that of other models.
The central density of Model H3 and H6 are lower, indicating that 
the systems are stretched out.
The central densities of other models are in between.
 
Fig. 13(b) gives the percentage of accumulated stellar mass as a function 
of $r$ for each considered model.
For the region with $r<3$, the solid curve (Model H1)
is at the top and the short-dashed curve (Model H3) is at the bottom. 
However, near $r=25$, all the curves approach to a similar value. 
Fig. 13(c)-(d) are the corresponding plots for the dark matter, and
the deviations between curves are smaller than those in Fig. 13(a)-(b).


In Table 2, we list
the percentages of stellar GUP (the first row) and
dark-matter GUP (the second row) at $t=39 T_{dyn}$ in hyperbolic mergers.
For the stellar GUP,
the average percentage is $m_{Hs}=0.3183$, and the standard deviation is 
$\sigma_{Hs}=0.0809$. The first interval 
$[m_{Hs}-\sigma_{Hs}, m_{Hs}+\sigma_{Hs}]=[0.2374,0.3992]$,
and the second interval
$[m_{Hs}-2\sigma_{Hs}, m_{Hs}+2\sigma_{Hs}]=[0.1565,0.4801]$.
The stellar GUPs of Model H1, H5, H6 are within the first interval, and
the ones of Model H2, H3, H4 are within the second interval.   
For the dark-matter GUP, the mean is  $m_{Hd}=14.1517$
with a standard deviation $\sigma_{Hd}=0.1446$. The 
$1\sigma$ interval
$[m_{Hd}-\sigma_{Hd}, m_{Hd}+\sigma_{Hd}]=[14.0071, 14.2963]$
and
$2\sigma$ interval 
$[m_{Hd}-2\sigma_{Hd}, m_{Hd}+2\sigma_{Hd}]=[13.8625, 14.4409]$.
Thus, the values of Model H1, H2, H5, H6 are in $1\sigma$ interval,
and the ones of Model H3, H4 are within $2\sigma$ interval
approximately.

\begin{table}[h]
\begin{center}
\begin{tabular}{|l||c|c|c|c|c|c|} \hline
{Model} & H1 & H2 & H3 & H4 & H5 & H6\\
\hline
\hline
stellar GUP (\%) & 0.37 & 0.21 & 0.42 & 0.21 & 0.33 & 0.37\\ \hline
dark-matter GUP (\%) & 14.22 & 14.12 & 14.33 & 13.86 & 14.19 & 14.19 \\ \hline
\end{tabular}
\caption[The GUP of Model H1-6]
{The percentages of 
stellar GUP and dark-matter GUP
in hyperbolic mergers.
}
\end{center}
\end{table} 

\subsection{The Credibility Tests}

N-body simulations have been used as powerful tools to study
the dynamical evolution of galaxies, and thus have helped to 
gain many important implications in this field.
However, technically, the detail setting of models would in general
lead to different numerical results. In N-body simulations of galaxies, 
one would usually concern that how the results vary with the force 
softening parameter and total number of particles.
  

In order to address this point, 
we have done three additional testing simulations, which would be 
called Run A, B, C.  
These three have different softening parameters or total number of particles.
For the simulations in Model P1-P6 and H1-H6, 
the total number of particles used 
in one merging event (consisting two galaxies) is 136000, the force softening
parameters of both stellar disc and dark halo are set to be 0.075 kpc.
We now define $N_M=136000$, $S_M=0.075$ kpc. All the details of testing
simulations are the same as in Model P1 except the values of 
total number of particles and softening parameters.

Table 3 shows the details of testing simulations, i.e. Run A, B, C, where
$N_T$ means the total number of particles used in one merging event,
$S_{h}$ means the softening parameter of dark halo, and
$S_{d}$ means the softening parameter of disc component.\\

\begin{table}[h]
\begin{center}
\begin{tabular}{|l||c|c|c|c|c|c|} \hline
& \multicolumn{1}{c}{$N_T$} \vline & \multicolumn{1}{c}{$S_{h}$} \vline & 
\multicolumn{1}{c}{ $S_{d}$ } \vline & 
$\delta E/E$ (\%)& Stellar GUP (\%) & Dark GUP (\%) \\
\hline
\hline
Run A & $1 N_M$ & $5 S_M$ & $5 S_M$ & 1.415 &  0.195 &  7.81 \\ \hline
Run B & $1 N_M$ & $5 S_M$ & $1 S_M$ &  1.375 &  0.205 &  7.87 \\ \hline 
Run C & $5 N_M$ & $1 S_M$ & $1 S_M$ & 1.160   &  0.180   & 7.98      \\ \hline 
Model P1 & $1 N_M$& $1 S_M$ & $1 S_M$ &  0.800 & 0.160 & 7.84 \\ \hline
\end{tabular}
\caption[The Tests]{The testing simulations: Run A, B, C.}
\end{center}
\end{table} 

The main results of Run A, B, C are also listed in Table 3,
where $\delta E/E$ means the percentage of energy variation over whole 
merging event, up to $t=39 T_{dyn}$.
To be convenient, the corresponding details and results of Model P1 are also
listed at the bottom of Table 3.

At first, in Run A, the total number of particles is the same as in Model P1,
but the softening parameters, $S_{h}$ and $S_{d}$ are five times larger 
than the values used in Model P1.
This is to test the effect of larger softening parameters while we keep 
that the halo and disc have the same value, i.e. $S_{h}=S_{d}$.
We find that $\delta E/E$ is slightly larger, the stellar GUP 
increases 0.035 percent, and the dark GUP decreases 0.03 percent.

Secondly, in Run B, the total number of particle and the 
disc softening parameter are the same as in Model P1,
but the halo softening parameter is now set to be five times larger 
than the disc softening parameter.
This is to test the effect that the halo and disc components have 
different softening parameters.
We find that $\delta E/E$ increases a bit, the stellar GUP 
increases 0.045 percent, and the dark GUP also increases 0.03 percent,
comparing with the results of Model P1.

Finally, in Run C, both the halo and disc softening parameters are the same
as in Model P1, but the total number of particles is now five time 
of the one in Model P1.
This is to test the effect of total number of particles.
We find that $\delta E/E=1.160$, the stellar GUP 
increases 0.02 percent, and the dark GUP also increases 0.14 percent,
comparing with the results of Model P1.

To summarize, these testing simulations show that 
the numerical variation of stellar GUP ranges 
from 0.02 to 0.045 percent, and the one of dark GUP
ranges from -0.03 to 0.14 percent due to the effects of 
softening parameters and total number of particles.
Comparing with the values in Model P1,
the fractions of variations range from 0.02/0.160 to 0.045/0.160, i.e. 
0.125 to 0.281 for the stellar GUP, and from -0.03/7.84 to 0.14/7.84,
i.e. -0.00383 to 0.01786 for the dark GUP.
The larger uncertainty for stellar GUP is due to the small
number of stellar GUP particles. 



\section{Concluding Remarks}



In order to investigate the relationship between
the galactic mergers and GUP,
we have used
the combined disc-halo systems to model the head-on merging events
of spiral galaxies. 
We consider both the parabolic and hyperbolic mergers, and we propose 
six models
for each of the mergers, 
with different relative orientations between two merging galaxies.

Our results show that the timing of the merging process
does not depend on the relative orientation.
Nevertheless,
the resulting structures of stellar components  
show a great variety for different disc orientations.

In terms of the production of GUP, we found that head-on merging
events produce much more dark-matter GUP than the stellar GUP.
This could be due to that the stellar particles are initially 
located at the very central part of whole systems, and thus
more difficult to become gravitational unbound.
On the other hand, the dark-matter particles distribute over a large
range, and those initially located at the outer part could be striped
out of the system easily.

Moreover, due to the higher energy injection rate of the hyperbolic merging
process, it
produces more GUP than the parabolic one (about a factor of two).
To understand the production of GUP at different times,
in Fig. 14(a)-(d), we plot the percentage of 
stellar GUP as a function of time.
It shows that, in Model P1-P6, the merging event 
starts to produce GUP quickly
at $t= 8 T_{dyn}$, 
which is actually the time when the first encounter happens.
After $t= 10 T_{dyn}$, very little stellar GUP particles are produced.  
In Model H1-H6, the production starts at $t= 5 T_{dyn}$. The production rate 
is very high from $t= 5 T_{dyn}$ to $t= 6 T_{dyn}$. Then, the stellar GUP 
slowly increases until $t= 15 T_{dyn}$.
Similarly, 
the corresponding plots for the dark-matter GUP are shown in Fig. 14(c)-(d).
Therefore,
the energy injection rate is essential 
for the production of GUPs.
In order to visualize these GUPs, the final distributions of stellar GUP 
were shown in Fig. 3 and Fig. 9
and dark-matter GUP particles are shown in Fig. 5 and Fig. 11, 
respectively.

On the other hand, 
we also calculated the mean and standard deviation of  
GUP percentages.  For parabolic mergers, the GUP  percentages
are all within or close to the boundary of $1\sigma$ interval.
For hyperbolic mergers, some model's GUP percentages
are in $1\sigma$ interval but some others are inside or near 
the boundary of $2\sigma$ interval.
Therefore, in addition to producing more GUP, 
the hyperbolic mergers also show larger scattering  
in terms of GUP production.

In general, 
our results show that, depending on the relative orientation
and the relative velocity, 
a head-on collision of a galaxy pair (with dark matter) 
would make less than one percent
of the initial stellar mass become the luminous GUP but 
about eight to fourteen percent of the dark matter become the dark GUP. 
Therefore, multiple mergers are needed to produce much more stellar
GUPs.

\section*{Acknowledgment}
We owe a debt of thanks to the anonymous referee for good suggestions
that improved the paper enormously.
We also thank the National Center for High-performance Computing
for computer time and facilities. 
This work is supported in part 
by the National Science Council, Taiwan, under 
NSC 97-2112-M-007-005.

\begin{figure}
\plotone{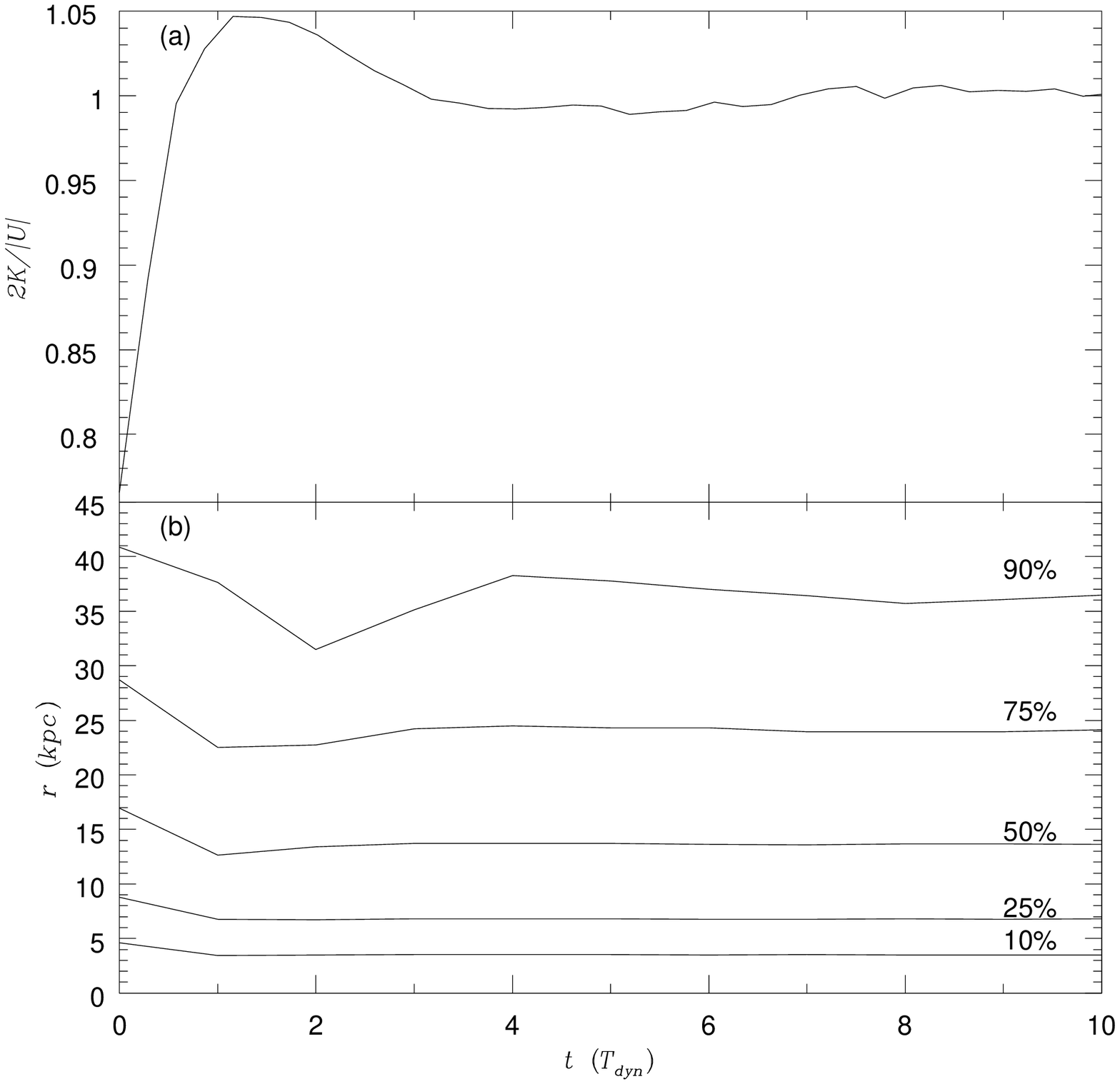}
\caption{The virial ratio $2K/|U|$ (top panel) and 
Lagrangian radii (bottom panel) of the disc-halo system 
as a function of time $t$, where the unit of $t$ is the dynamical time
$T_{dyn}$} 
\end{figure}

\clearpage
\begin{figure}
\plotone{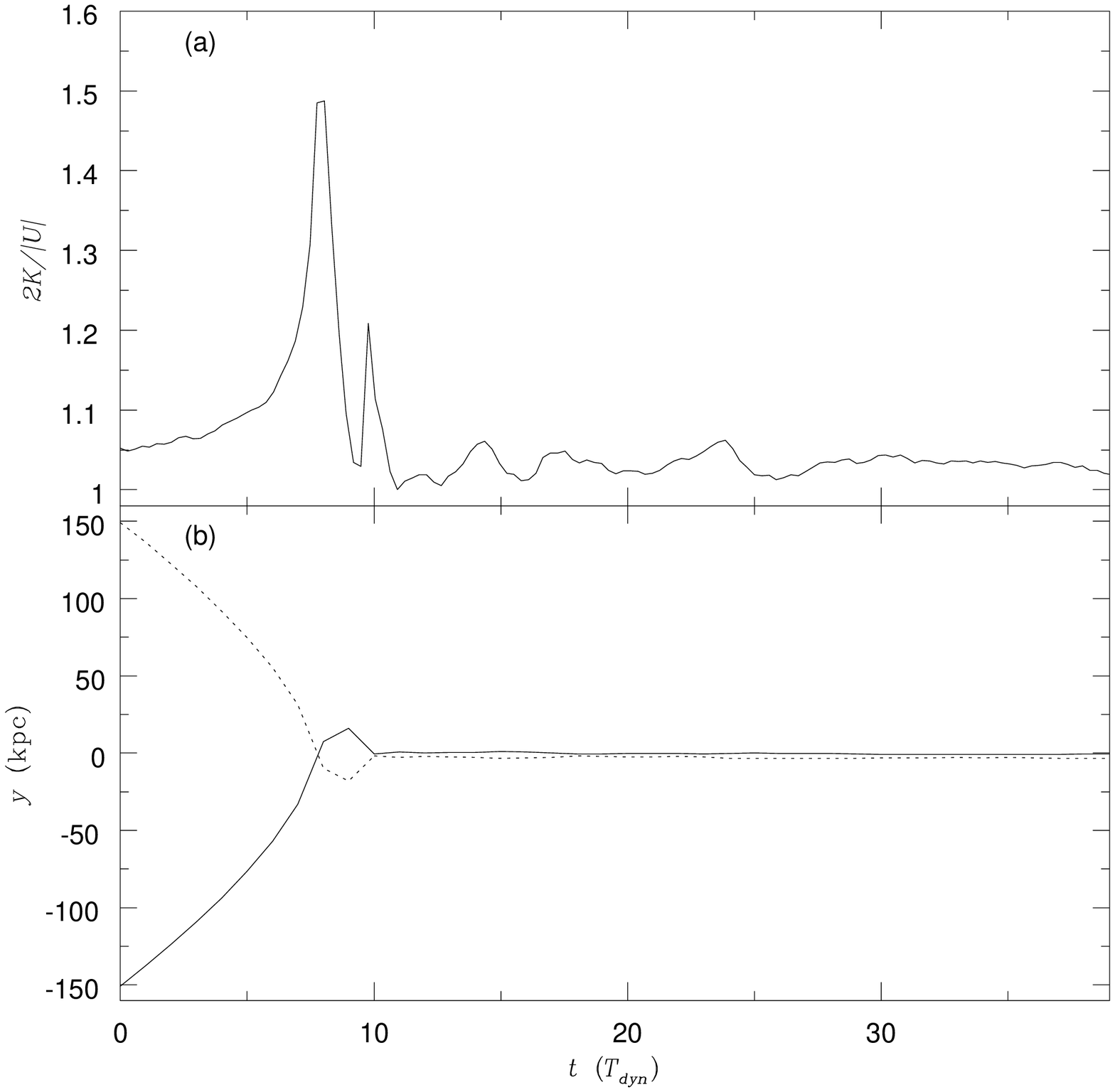}
 \caption{The evolution of Model P1. 
(a) The virial ratio $2K/|U|$ of the merger 
as a function of time $t$.
(b) The distance between two stellar components as a function of time 
during the merging process, where the solid and dotted curves
are for the locations of two stellar components. 
}
\end{figure}

\clearpage
\begin{figure}
\rotatebox{270}{
\epsscale{0.7}
\plotone{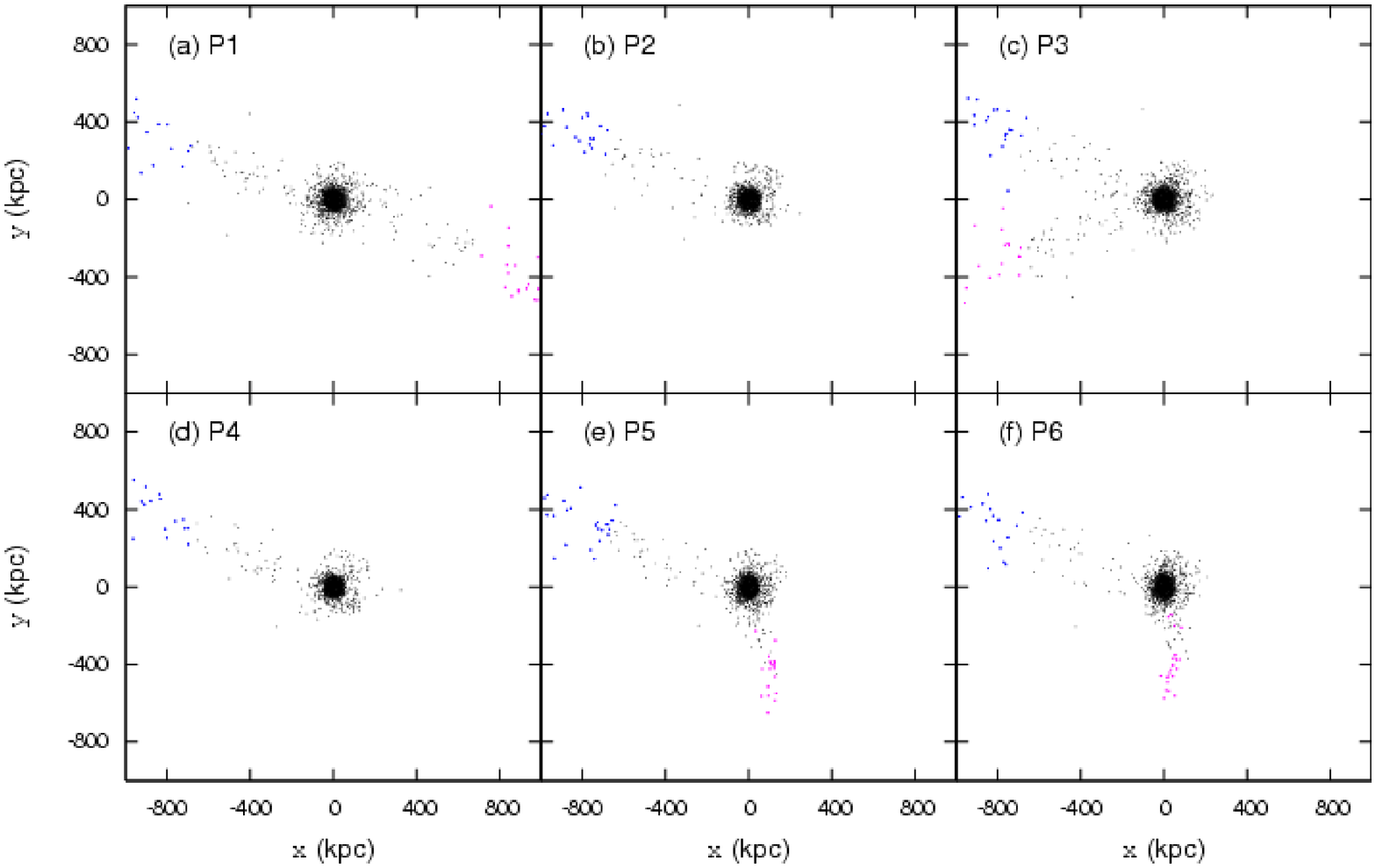}
}
\caption{The distributions of stellar particles on 
the $x-y$ plane at $t=39 T_{dyn}$ in Model P1-P6.
The blue dots represent the unbound particles which
were belong to the galaxy initially located at $(x,y,z)=(0, -150, 0)$,
and the pink dots represent the unbound particles which
were belong to another galaxy.
}
\end{figure}

\clearpage
\begin{figure}
\rotatebox{270}{
\epsscale{0.8}
\plotone{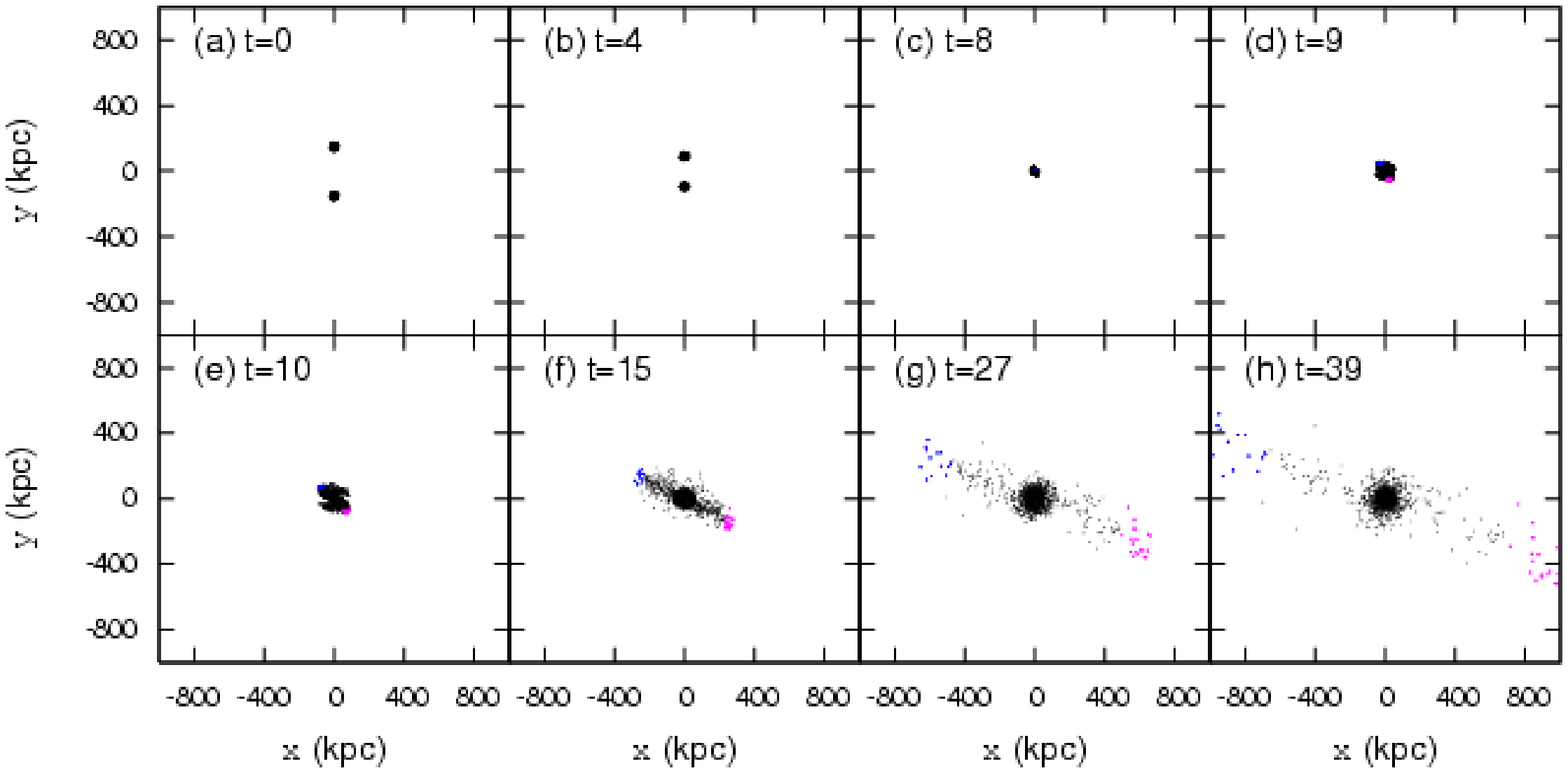}
}
\caption{The distributions of stellar particles on 
the $x-y$ plane in Model P1.
(a) t=0; (b) t=4 $T_{dyn}$; (c) t=8 $T_{dyn}$; (d) t=9 $T_{dyn}$;
(e) t=10 $T_{dyn}$; (f) t=15 $T_{dyn}$; (g) t=27 $T_{dyn}$; 
(h) t=39 $T_{dyn}$.
The blue dots represent the unbound particles which
were belong to the galaxy initially located at $(x,y,z)=(0, -150, 0)$,
and the pink dots represent the unbound particles which
were belong to another galaxy.
}
\end{figure}

\clearpage
\begin{figure}
\rotatebox{270}{
\epsscale{0.7}
\plotone{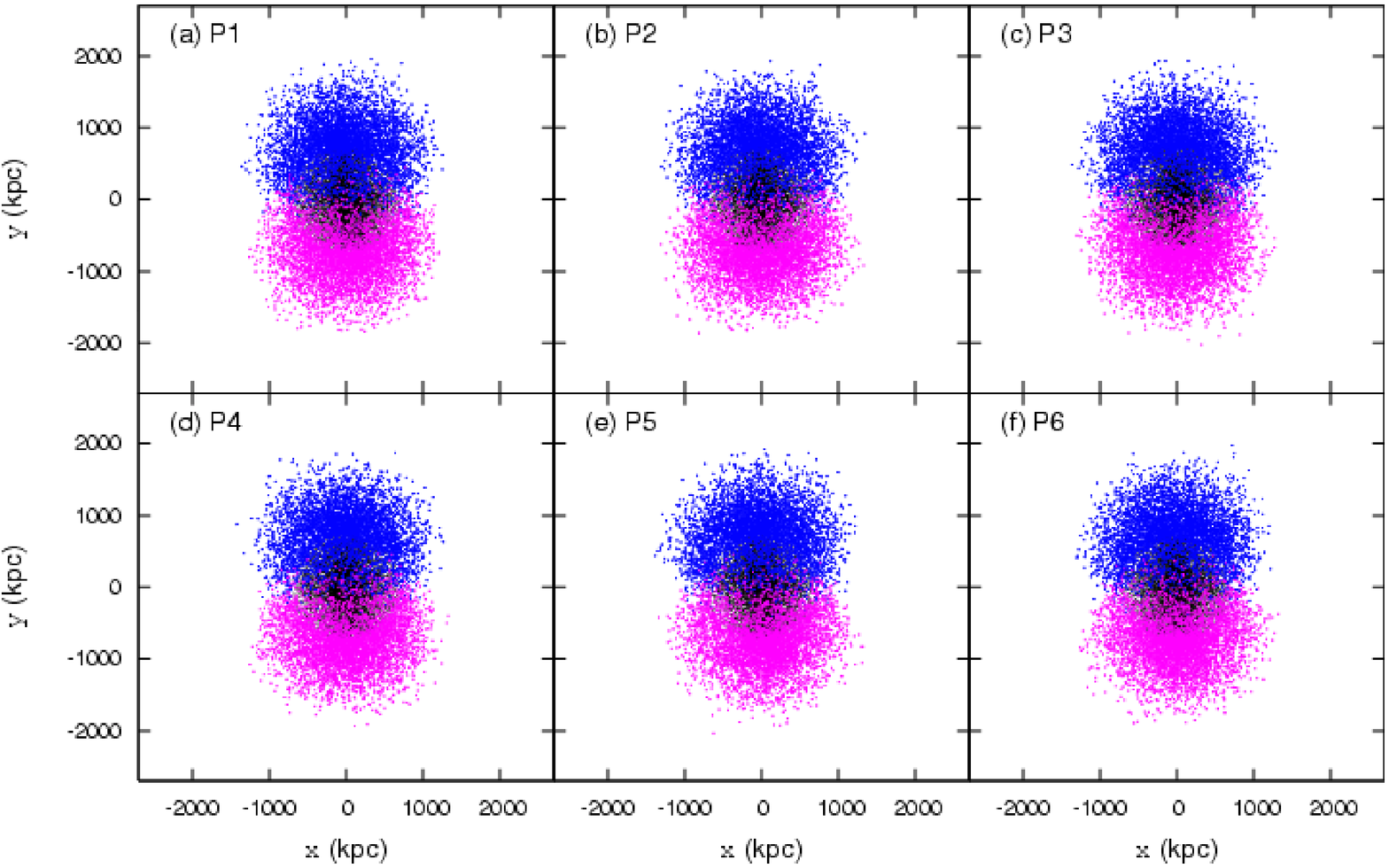}
}
\caption{The distributions of halo particles on 
the $x-y$ plane  at $t=39 T_{dyn}$ in Model P1-P6.
The blue dots represent the unbound particles which
were belong to the galaxy initially located at $(x,y,z)=(0, -150, 0)$,
and the pink dots represent the unbound particles which
were belong to another galaxy.
}
\end{figure}

\clearpage
\begin{figure}
\rotatebox{270}{
\epsscale{0.8}
\plotone{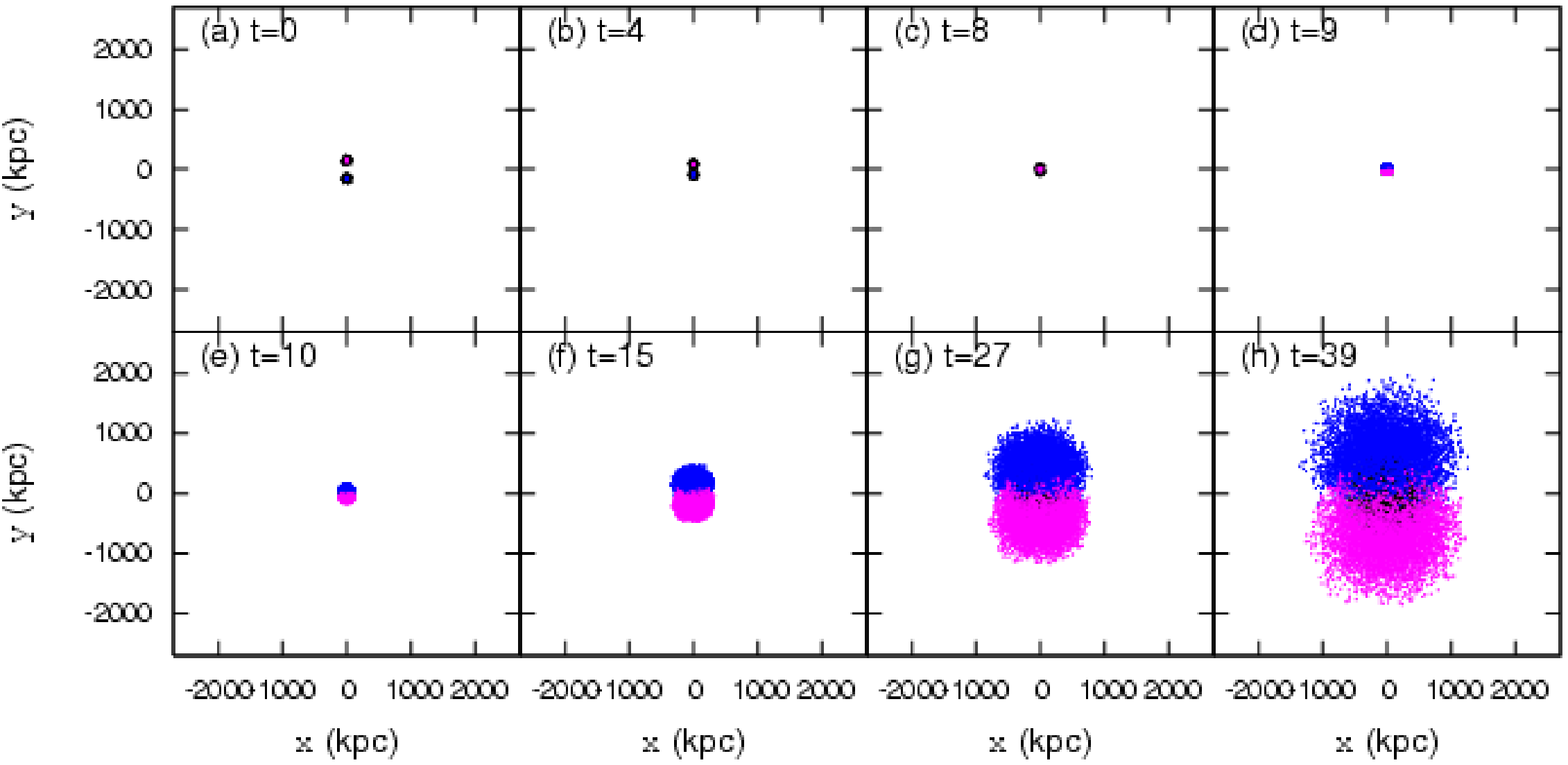}
}
\caption{The distributions of halo particles on 
the $x-y$ plane in Model P1.
(a) t=0; (b) t=4 $T_{dyn}$; (c) t=8 $T_{dyn}$; (d) t=9 $T_{dyn}$;
(e) t=10 $T_{dyn}$; (f) t=15 $T_{dyn}$; (g) t=27 $T_{dyn}$; 
(h) t=39 $T_{dyn}$.
The blue dots represent the unbound particles which
were belong to the galaxy initially located at $(x,y,z)=(0, -150, 0)$,
and the pink dots represent the unbound particles which
were belong to another galaxy.
}
\end{figure}

\clearpage
\begin{figure}
\epsscale{1.0}
\plotone{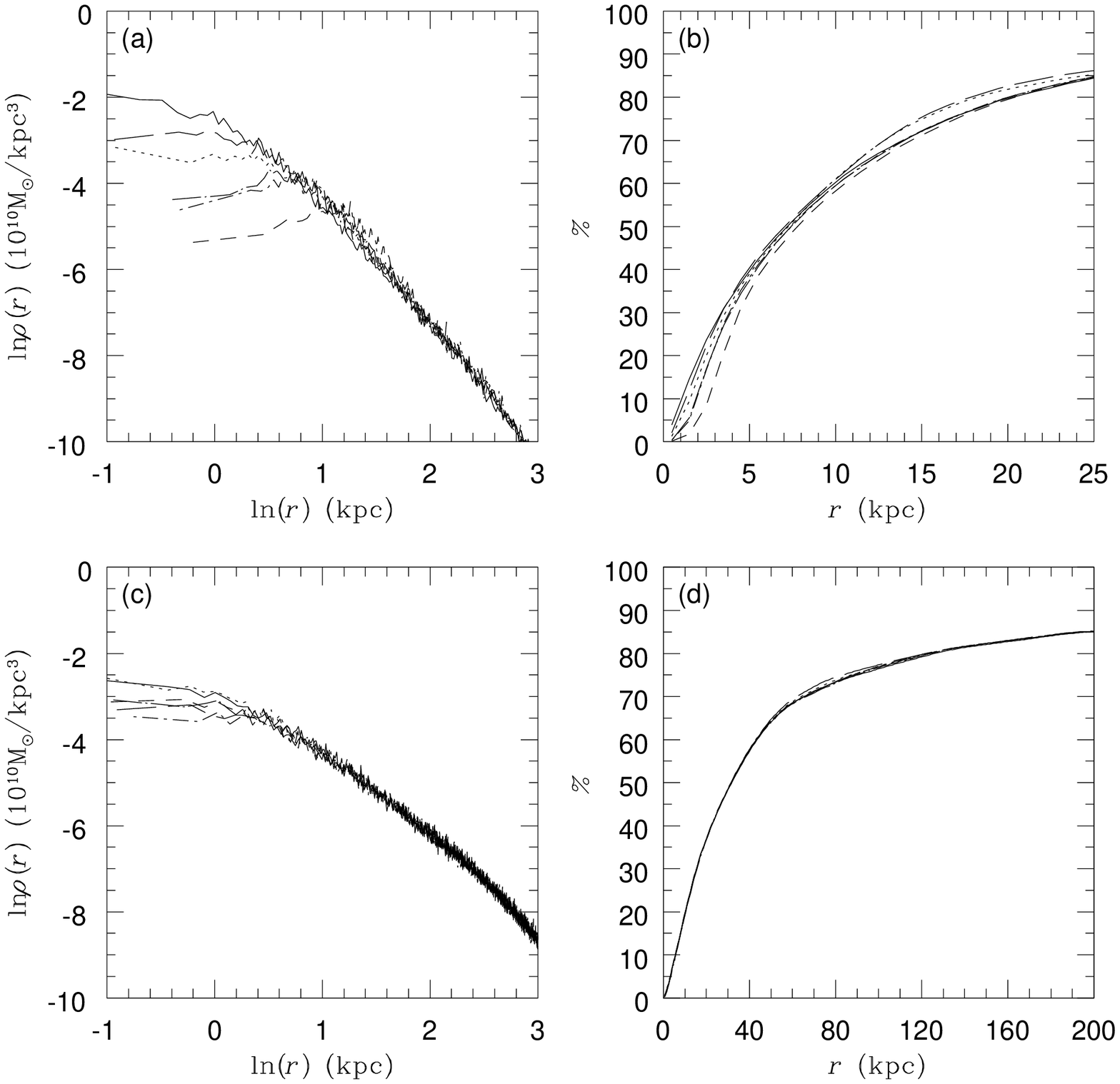}
\caption{The profiles of parabolic mergers at $t= 39 T_{dyn}$.
(a) The density profiles of stellar components.
(b) The percentage of stellar mass as a function of $r$.
(c) The density profiles of dark-matter components.
(d) The percentage of dark-matter mass as a function of $r$.
The solid, dotted, short-dashed, long-dashed,  dotted short-dashed, and
dotted long-dashed curve  
is for Model P1-P6, respectively. The unit of $r$ is 
kpc and the unit of $\rho(r)$
is $10^{10}M_{\bigodot}/ \rm{kpc^3}$.
}
\end{figure}

\clearpage
\begin{figure}
\epsscale{1.0}
\plotone{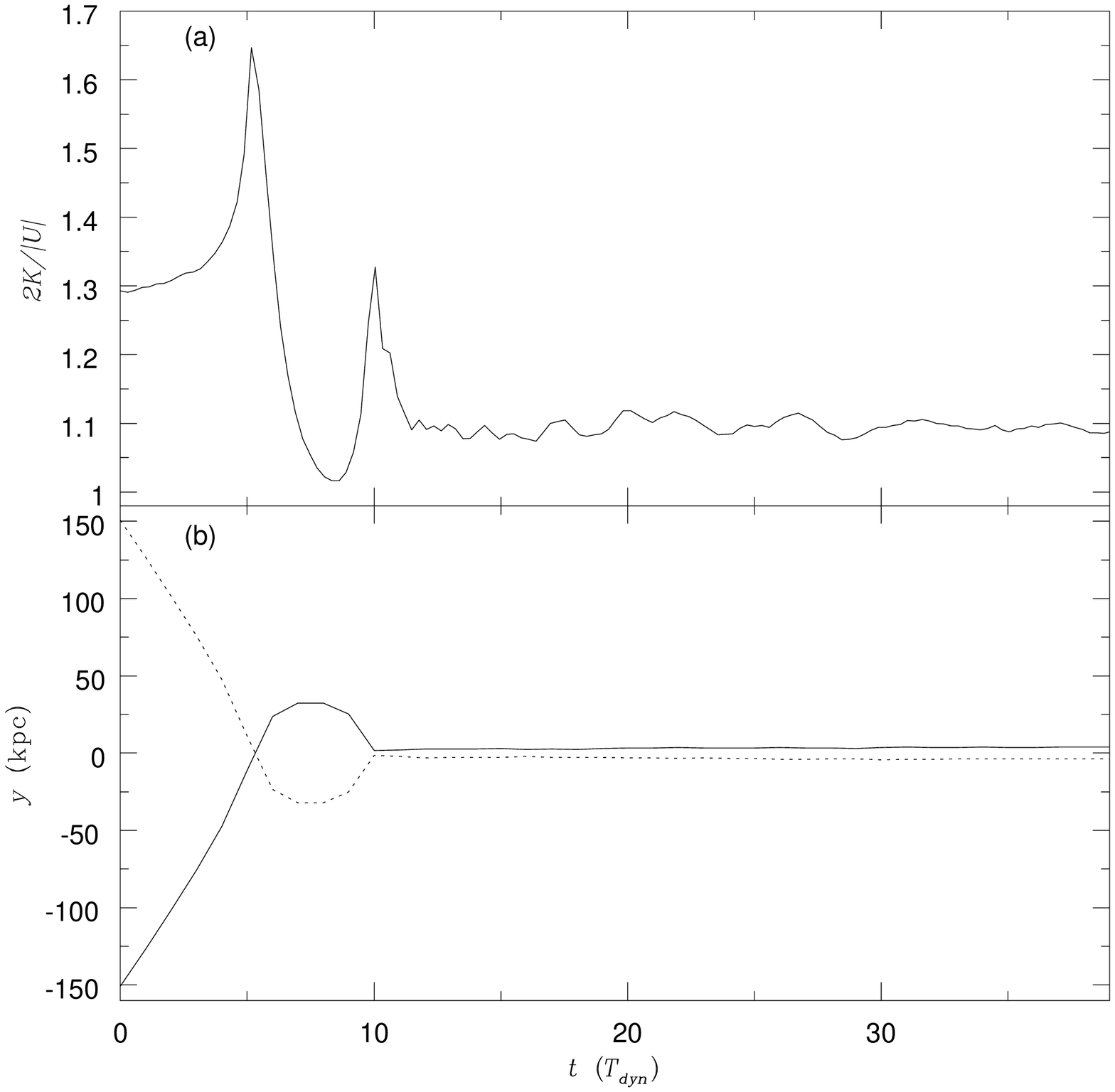}
 \caption{The evolution of Model H3. 
(a) The virial ratio $2K/|U|$ of the merger 
as a function of time $t$.
(b) The distance between two stellar components as a function of time 
during the merging process, where the solid and dotted curves
are for the locations of two stellar components. 
}
\end{figure}

\clearpage
\begin{figure}
\rotatebox{270}{
\epsscale{0.7}
\plotone{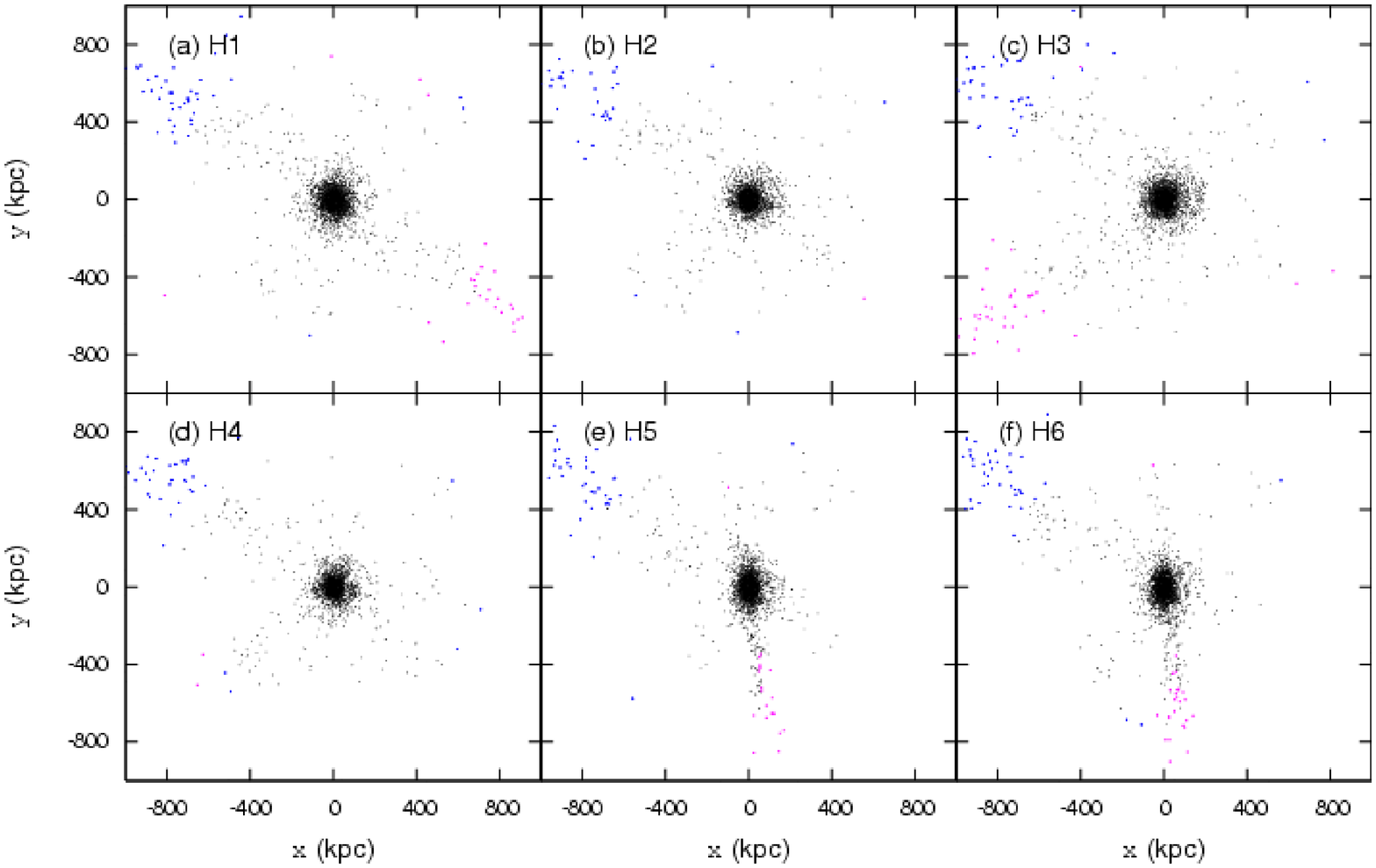}
}
\caption{The distributions of stellar particles on 
the $x-y$ plane at $t=39 T_{dyn}$ in Model H1-H6.
The blue dots represent the unbound particles which
were belong to the galaxy initially located at $(x,y,z)=(0, -150, 0)$,
and the pink dots represent the unbound particles which
were belong to another galaxy.
}
\end{figure}

\clearpage
\begin{figure}
\rotatebox{270}{
\epsscale{0.8}
\plotone{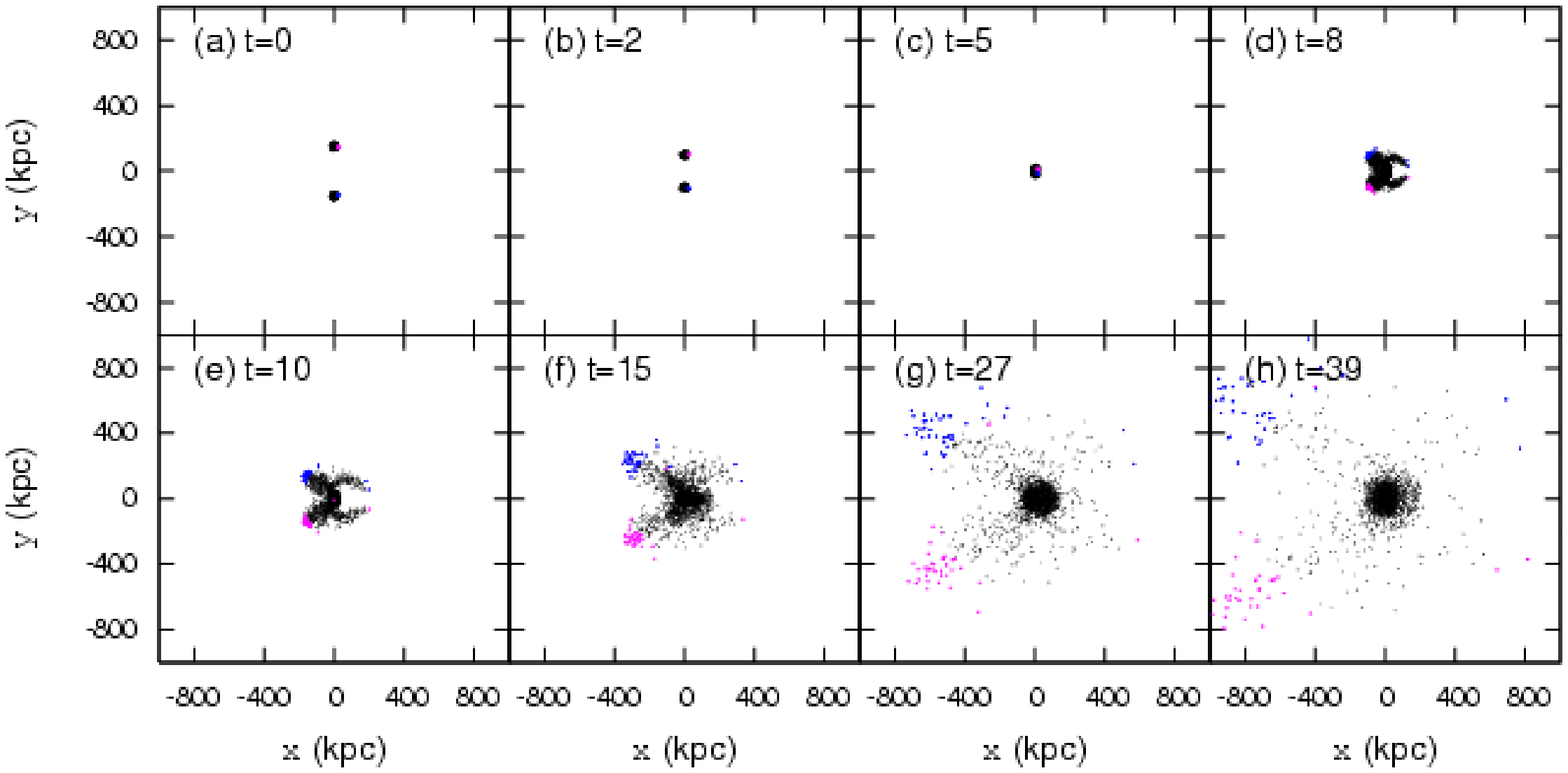}
}
\caption{The distributions of stellar particles on 
the $x-y$ plane in Model H3.
(a) t=0; (b) t=2 $T_{dyn}$; (c) t=5 $T_{dyn}$; (d) t=8 $T_{dyn}$;
(e) t=10 $T_{dyn}$; (f) t=15 $T_{dyn}$; (g) t=27 $T_{dyn}$; 
(h) t=39 $T_{dyn}$.
The blue dots represent the unbound particles which
were belong to the galaxy initially located at $(x,y,z)=(0, -150, 0)$,
and the pink dots represent the unbound particles which
were belong to another galaxy.
}
\end{figure}

\clearpage
\begin{figure}
\rotatebox{270}{
\epsscale{0.7}
\plotone{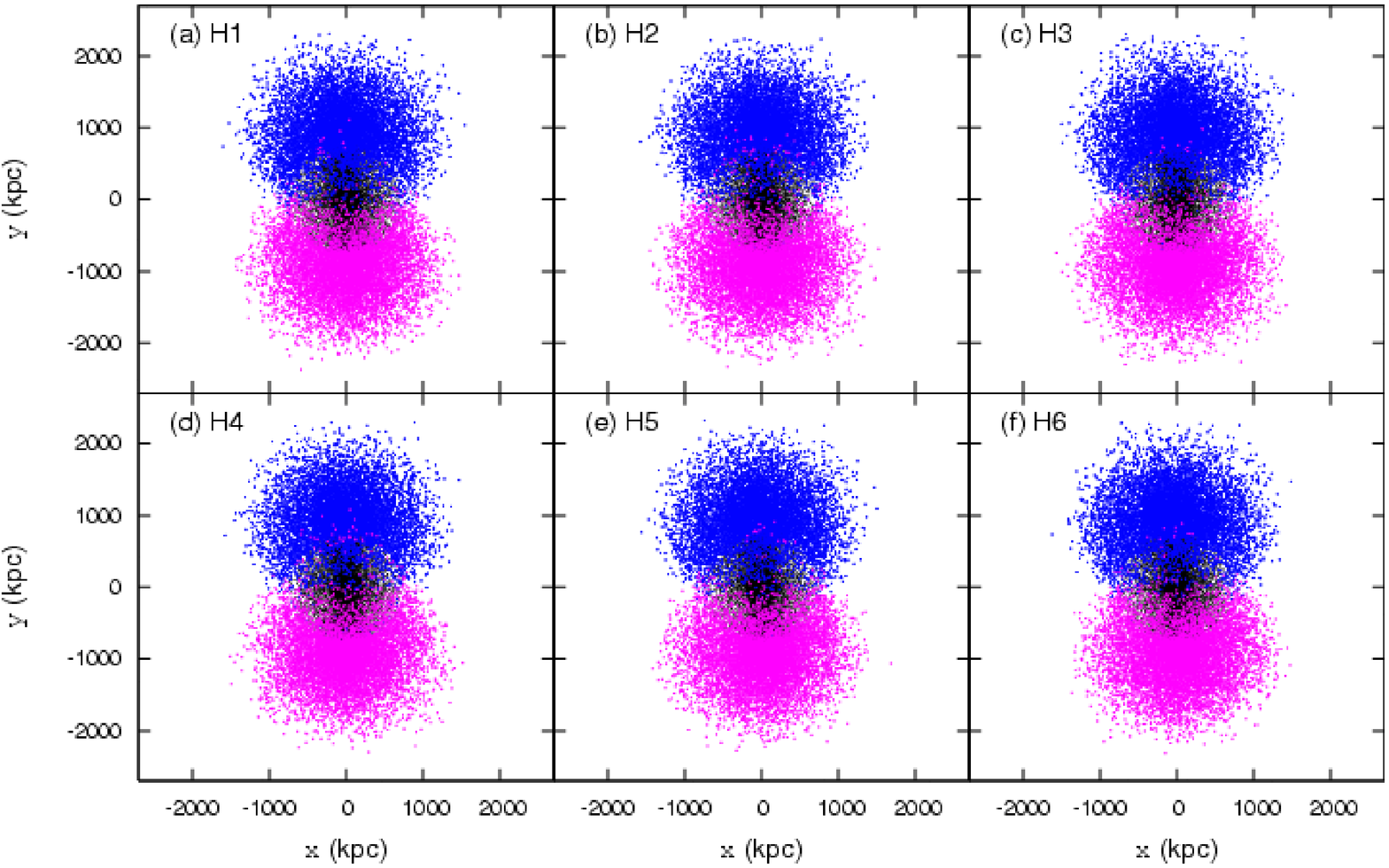}
}
\caption{The distributions of halo particles on 
the $x-y$ plane at $t=39 T_{dyn}$ in Model H1-H6.
The blue dots represent the unbound particles which
were belong to the galaxy initially located at $(x,y,z)=(0, -150, 0)$,
and the pink dots represent the unbound particles which
were belong to another galaxy.
}
\end{figure}

\clearpage
\begin{figure}
\rotatebox{270}{
\epsscale{0.8}
\plotone{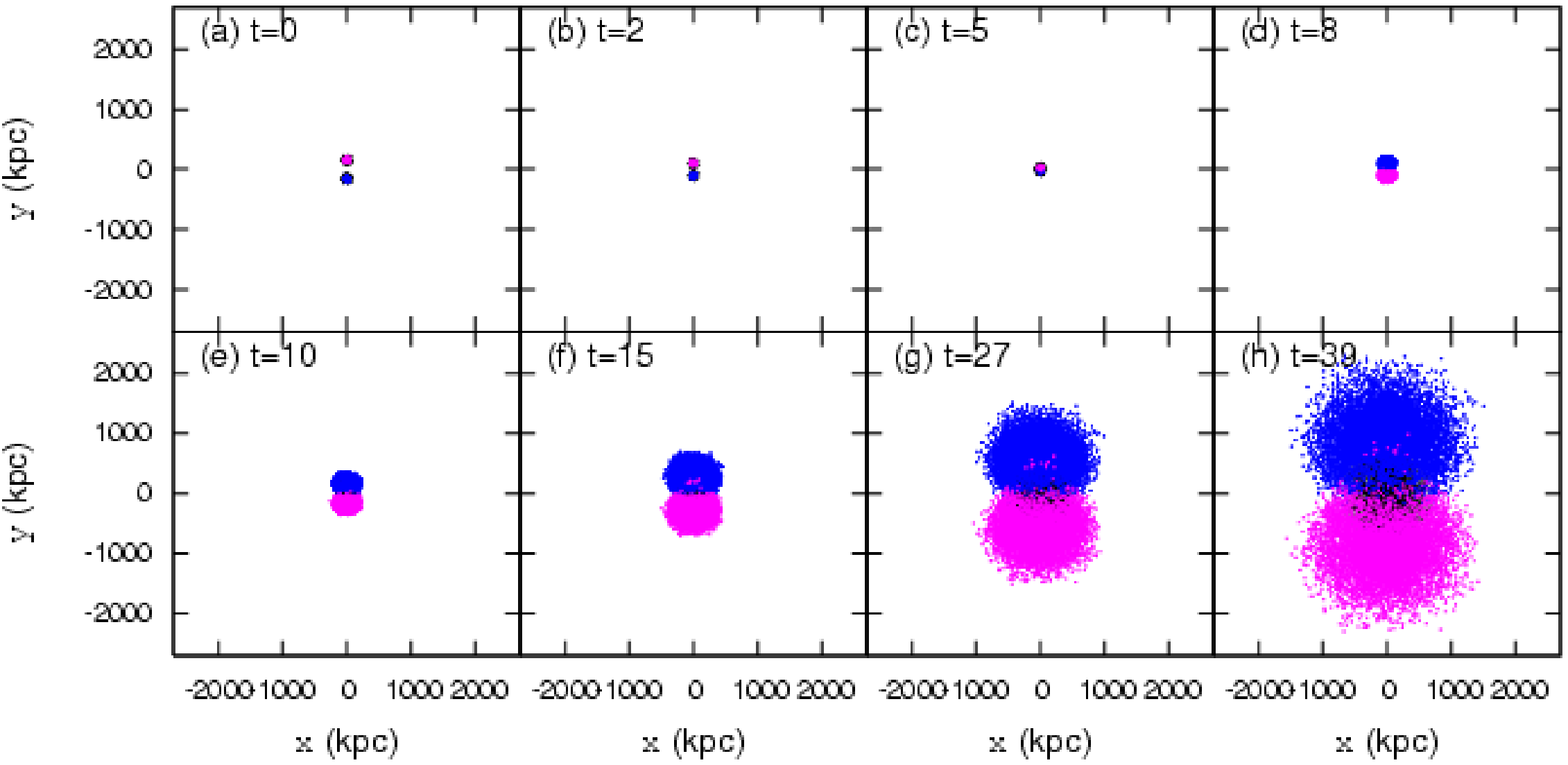}
}
\caption{The distributions of halo particles on 
the $x-y$ plane in Model H3.
(a) t=0; (b) t=2 $T_{dyn}$; (c) t=5 $T_{dyn}$; (d) t=8 $T_{dyn}$;
(e) t=10 $T_{dyn}$; (f) t=15 $T_{dyn}$; (g) t=27 $T_{dyn}$; 
(h) t=39 $T_{dyn}$.
The blue dots represent the unbound particles which
were belong to the galaxy initially located at $(x,y,z)=(0, -150, 0)$,
and the pink dots represent the unbound particles which
were belong to another galaxy.
}
\end{figure}

\clearpage
\begin{figure}
\epsscale{1.0}
\plotone{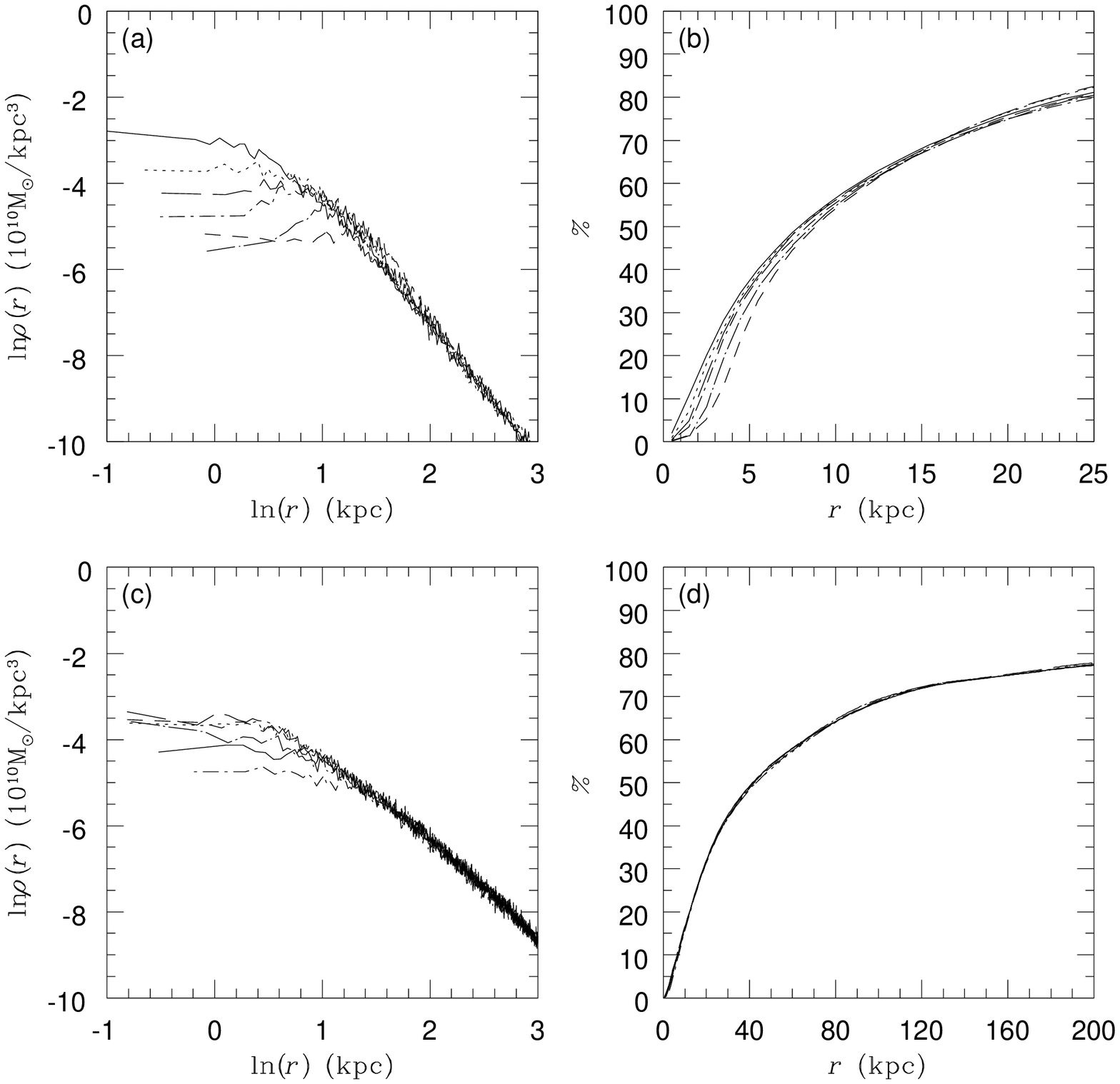}
\caption{The profiles of hyperbolic mergers at $t= 39 T_{dyn}$.
(a) The density profiles of stellar components.
(b) The percentage of stellar mass as a function of $r$.
(c) The density profiles of dark-matter components.
(d) The percentage of dark-matter mass as a function of $r$.
The solid, dotted, short-dashed, long-dashed,  dotted short-dashed, and
dotted long-dashed curve  
is for Model H1-H6, respectively. The unit of $r$ is 
kpc and the unit of $\rho(r)$
is $10^{10}M_{\bigodot}/ \rm{kpc^3}$.
}
\end{figure}

\clearpage

\begin{figure}
\epsscale{1.0}
\plotone{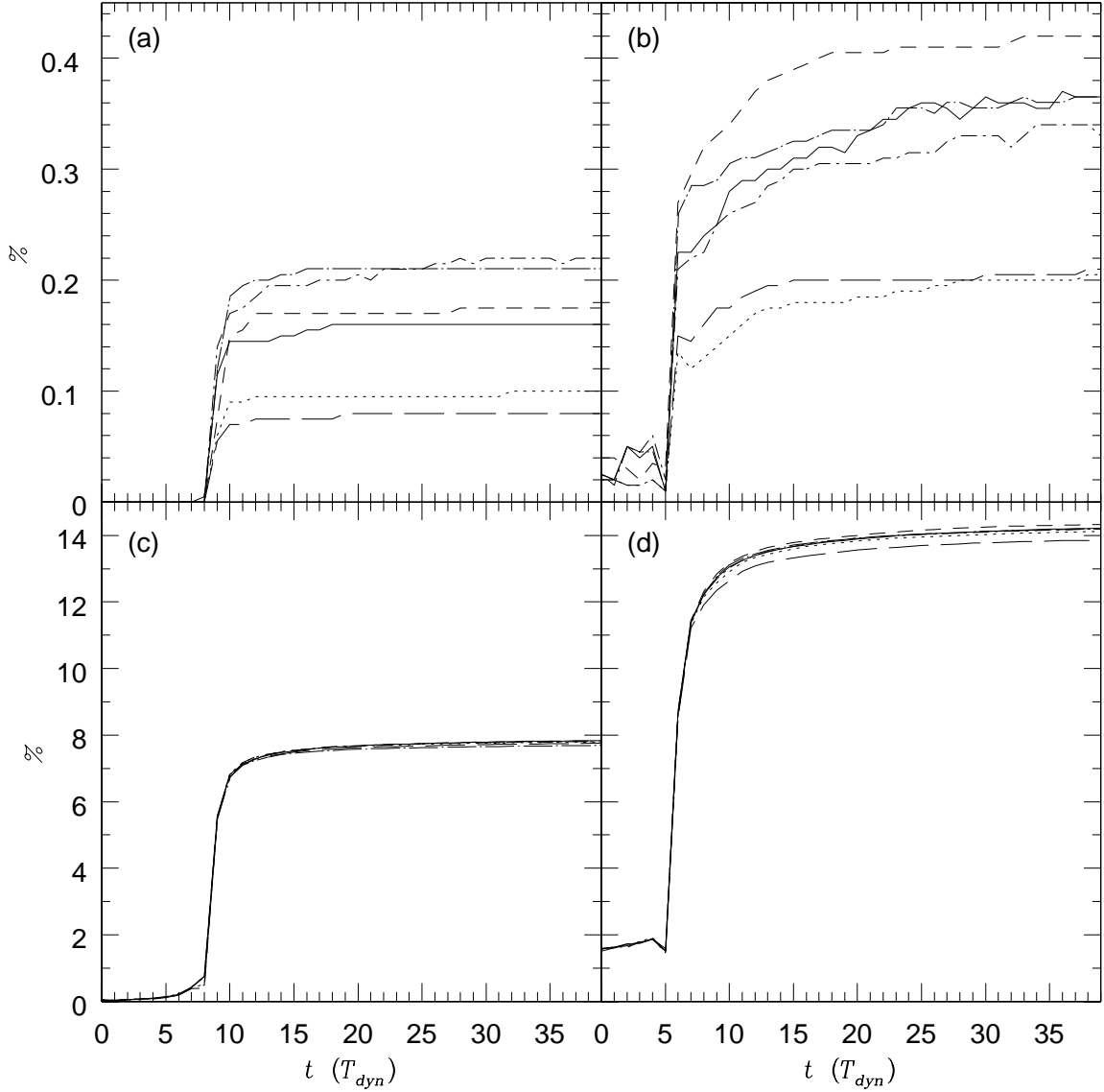}
\caption{
(a) The percentage of stellar GUP as a function of time in Model P1-P6.
(b) The percentage of stellar GUP as a function of time in Model H1-H6.
(c) The percentage of dark GUP as a function of time in Model P1-P6.
(d) The percentage of dark GUP as a function of time in Model H1-H6.
The solid, dotted, short-dashed, long-dashed,  dotted short-dashed, and
dotted long-dashed curve  
is for Model P1-P6 (H1-H6), respectively.
}
\end{figure}

\end{document}